\documentclass[10pt,twocolumn,twoside]{IEEEtran} 
\ifCLASSINFOpdf
   \usepackage[pdftex]{graphicx}
\else
   \usepackage[dvips]{graphicx}
\fi
\usepackage[cmex10]{amsmath}
\usepackage{amsfonts,amssymb}
\usepackage{color}

\newcommand{\vc}[1]{{\boldsymbol{#1}}}

\newtheorem{theorem}{Theorem}
\newtheorem{proposition}{Proposition}

\newtheorem{definition}{Definition}

\begin{document}
%
\title{Discrete Signal Reconstruction by\\ Sum of Absolute Values}
%
%
%

\author{Masaaki~Nagahara,~\IEEEmembership{Senior Member,~IEEE,}%
\thanks{Copyright (c) 2015 IEEE. Personal use of this material is permitted. However, permission to use this material for any other purposes must be obtained from the IEEE by sending a request to pubs-permissions@ieee.org.}        
\thanks{M. Nagahara is with
	Graduate School of Informatics,
	Kyoto University, Kyoto, 606-8501, Japan;
	email: nagahara@ieee.org (corresponding author)}
}

%
%

\markboth{Journal of \LaTeX\ Class Files,~Vol.~11, No.~4, December~2012}%
{Shell \MakeLowercase{\textit{et al.}}: Bare Demo of IEEEtran.cls for Journals}
%



\maketitle

\begin{abstract}
In this letter, we consider a problem of reconstructing
an unknown discrete signal taking values in a finite alphabet
from incomplete linear measurements.
The difficulty of this problem is that 
the computational complexity of the reconstruction is exponential as it is.
To overcome this difficulty, we extend the idea of compressed sensing, and propose
to solve the problem by minimizing the sum of weighted absolute values.
We assume that the probability distribution defined on an alphabet is known,
and formulate the reconstruction problem as linear programming.
Examples are shown to illustrate that the proposed method is effective.
\end{abstract}


%
\IEEEpeerreviewmaketitle

\section{Introduction}
\label{sec:introduction}
%
%
%
%


Signal reconstruction is a fundamental problem in signal processing.
Recently, a paradigm called {\em compressed sensing}~\cite{Don06,EldKut,HayNagTan13}
has been proposed for signal reconstruction from incomplete measurements.
The idea of compressed sensing is to utilize the property of {\em sparsity}
in the original signal;
if the original signal is sufficiently sparse, practical algorithms
such as the basis pursuit~\cite{CheDonSau98}, the orthogonal matching pursuit~\cite{PatRezKri93}, etc,
may give exact recovery under an assumption on the measurement matrix
(see e.g. \cite{HayNagTan13}).

On the other hand, it is also important to reconstruct discrete signals
whose elements are generated from a finite alphabet
with a known probability distribution.
This type of reconstruction, called {\em discrete signal reconstruction},
arises in black-and-white or grayscale sparse image reconstruction~\cite{Dua+08,BioColMag14},
blind estimation in digital communications~\cite{VeeTalPau95},
machine-type multi-user communications \cite{Kno+14},
discrete control signal design~\cite{BemMor99}, to name a few.

The difficulty of discrete signal reconstruction is that
the reconstruction has a combinatorial nature
and the computational time becomes exponential.
For example, 200-dimensional signal reconstruction
with two symbols (i.e. binary signal reconstruction)
needs at worst about $1.5\times 10^{36}$ years
with a computer of 34 peta FLOPS (see Section \ref{sec:problem} below),
which cannot be executed, obviously.

To overcome this difficulty, we borrow the idea of compressed sensing
based on $\ell^1$ optimization as used in the basis pursuit
\cite{CheDonSau98}.
Our idea is that if the original discrete signal, say $\vc{x}$, 
includes $L$ symbols $r_1,r_2,\dots,r_L$, then
each vector $\vc{x}-r_i$ is sparse.
For example, a binary vector $\vc{x}$
on alphabet $\{1,-1\}$
includes a number of $1$ and $-1$, and hence
both $\vc{x}-1$ and $\vc{x}+1$ are sparse.
To recover such a discrete signal,
we propose to minimize the {\em sum of weighted absolute values}
of the elements in $\vc{x}-r_i$.
The weights are determined by the probability distribution on the alphabet.
The problem is reduced to a standard linear programming problem,
and effectively solved by numerical softwares such as {\tt cvx} in 
{\tt MATLAB}~\cite{cvx,GraBoy08}.

For discrete signal reconstruction, there have been researches
based on compressed sensing,
called \emph{integer compressed sensing}:
\cite{TiaLeuLot09} has proposed a Bayesian-based method,
which works only for binary sparse signals (i.e., $0$-$1$ valued signals that contain
many 0's).
\cite{Kno+14} considers arbitrary finite alphabet that contains $0$.
More recently, motivated by decision feedback equalization,
\cite{IliStr12} proposes to use $\ell^1$ optimization for discrete signal estimation
under the assumption of sparsity.
\cite{SarBarBar06,DasVis13,BioColMag14} also propose methods based on
the finiteness of the measurement matrix (i.e., the elements of the measurement matrix
are also in a finite alphabet).
As mentioned in \cite{SarBarBar06}, this type of integer compressed sensing is
connected with error correcting coding.
Compared with these researches, the proposed method in this paper
considers arbitrary finite alphabet that does not necessarily contain $0$.

The remainder of this letter is organized as follows:
Section \ref{sec:problem} formulates the problem of
discrete signal reconstruction and discusses the difficulty
of the problem.
Section \ref{sec:solution} proposes to use the sum
of weighted absolute values for discrete signal reconstruction, and 
show a sufficient and necessary condition for exact recovery
by extending the notion of the null space property
\cite{CohDahDeV09}.
Examples of one-dimensional signals and two-dimensional images
are included in Section \ref{sec:example}.
Section \ref{sec:conclusion} draws conclusions.

\subsection*{Notation}
For a vector $\vc{x}=[x_1,x_2,\ldots,x_N]^\top \in{\mathbb{R}}^N$,
we define the $\ell^1$ and $\ell^2$ norms respectively as
\[
 \|\vc{x}\|_1 \triangleq \sum_{n=1}^N |x_n|,~ \|\vc{x}\|_2 \triangleq \sqrt{\vc{x}^\top\vc{x}},
\] 
where $\top$ denotes the transpose.
For a vector $\vc{x}$ and a scalar $r$, we define
\[
 \vc{x}-r \triangleq [x_1-r,x_2-r,\ldots,x_N-r]^\top.
\]
For a matrix $\Phi\in{\mathbb{C}}^{M\times N}$, 
$\mathrm{ker}\,\Phi$ is the kernel (or the null space) of $\Phi$, that is,
\[
 \mathrm{ker}\,\Phi \triangleq \{\vc{x}\in{\mathbb{C}}^N: \Phi\vc{x}=\vc{0}\}.
\] 
$I_n$ is the $n$-dimensional identity matrix, and $\vc{1}_k$ is a $k$-dimensional vector
whose elements are all $1$, that is, 
\[
 \vc{1}_k\triangleq[1,1,\ldots,1]^\top \in {\mathbb{R}}^k.
\] 
For two matrices $A\in{\mathbb{R}}^{M\times N}$ and 
$B\in{\mathbb{R}}^{K\times L}$, $A\otimes B$ denotes
the Kronecker product, that is,
\[
 A\otimes B \triangleq 
 \begin{bmatrix}
 A_{11}B & A_{12}B & \ldots & A_{1N}B\\
 \vdots & \vdots & \ddots & \vdots\\
 A_{M1}B & A_{M2}B & \ldots & A_{MN}B
 \end{bmatrix}
 \in{\mathbb{R}}^{M\!K\times N\!L},
\]
where $A_{ij}$ is the $ij$-th element of $A$.
For two real-valued vectors $\vc{x}$ and $\vc{y}$ of the same size,
$\vc{x}\leq\vc{y}$ denotes the element-wise inequality,
that is, $\vc{x}\leq\vc{y}$ means $x_i\leq y_i$ for all $i$.

\section{Problem Formulation}
\label{sec:problem}
Assume that the original signal $\vc{x}$ is an $N$-dimensional vector
whose elements are discrete. That is,
\[
 \begin{split}
 \vc{x} &\triangleq [x_1,x_2,\ldots,x_N]^\top \in {\mathcal X}^N,\\
 {\mathcal X} &\triangleq \{r_1,r_2,\ldots,r_L\},
 \end{split}
\]
where $r_i\in{\mathbb R}$ and we assume
\begin{equation}
r_1<r_2<\cdots<r_L.
\label{eq:rl}
\end{equation}
If a symbol, say $r$, is a complex number, then taking
$r_j \triangleq \mathrm{Re}(r)$ and $r_k \triangleq \mathrm{Im}(r)$
gives a real-valued alphabet,
and hence the assumption \eqref{eq:rl} is not restrictive.
We here assume that the values of $r_i$ are known and the probability distribution
of them is given by
\begin{equation}
 {\mathbb P}(r_i) = p_i,\quad i=1,2,\ldots,L,
 \label{eq:Prob}
\end{equation}
where we assume
\begin{equation}
 p_i >0,\quad p_1+p_2+\cdots+p_L=1.
 \label{eq:p}
\end{equation}

Then we consider a linear measurement process modelled by
\begin{equation}
 \vc{y} = \Phi\vc{x} \in {\mathbb C}^M,
 \label{eq:measurement}
\end{equation}
where $\Phi \in {\mathbb C}^{M\times N}$.
Note that we consider a complex-valued matrix $\Phi$
since $\Phi$ can be constructed from e.g. a complex-valued DFT
(Discrete Fourier Transform) matrix; see the image processing
example in Section \ref{sec:example}.
We assume incomplete measurements, that is,
$M < N$.
The objective here is to reconstruct $\vc{x}\in{\mathcal X}^N$ 
from the measurement vector $\vc{y}\in{\mathbb C}^M$ in \eqref{eq:measurement}.

First of all, we discuss the uniqueness of the solution of the discrete signal reconstruction.
We have the following proposition:
\begin{proposition}
\label{prop:uniqueness}
Given $\Phi\in{\mathbb{C}}^{M\times N}$, the following properties are equivalent:
\begin{enumerate}
 \item[(A)] 
 If $\Phi\vc{x}_1=\Phi\vc{x}_2$ and both $\vc{x}_1$ and $\vc{x}_2$ are in $\mathcal{X}^N$,
 then $\vc{x}_1=\vc{x}_2$.
 \item[(B)] Define the difference set of ${\mathcal{X}}$ as
 \[
  \tilde{{\mathcal{X}}} \triangleq \{r_i-r_j: i, j = 1,2,\ldots,L\}.
 \]
Then
\begin{equation}
 \mathrm{ker}\,\Phi \cap \tilde{\mathcal{X}}^N=\{\vc{0}\}.
 \label{eq:kerP}
\end{equation}
\item[(C)] The matrix $\Phi$ is injective as a map from $\tilde{\mathcal{X}}^N$ to $\mathbb{C}^M$.
\end{enumerate}
\end{proposition}
\begin{IEEEproof}
(A)$\Rightarrow$(B):~Assume (A) holds.
Take any $\vc{v}\in\mathrm{ker}\,\Phi\cap\tilde{\mathcal{X}}^N$.
Since $\vc{v}\in\tilde{\mathcal{X}}^N$, there exist
$\vc{x}_1, \vc{x}_2 \in {\mathcal{X}}^N$ such that $\vc{v}=\vc{x}_1-\vc{x}_2$.
Then we have
$\Phi\vc{v}=\Phi\vc{x}_1-\Phi\vc{x}_2=\vc{0}$
since $\vc{v}\in\mathrm{ker}\,\Phi$.
Then from (A), we have $\vc{x}_1=\vc{x}_2$. It follows that $\vc{v}=\vc{0}$.

(B)$\Rightarrow$(C):~Assume (B) holds.
Take any $\vc{v}\in\tilde{{\mathcal{X}}}^N$
and assume $\Phi\vc{v}=\vc{0}$.
Then from (B), we have $\vc{v}=\vc{0}$.
This proves $\Phi$ is an injective map 
from $\tilde{\mathcal{X}}^N$ to ${\mathbb{C}}^M$.

(C)$\Rightarrow$(A):~
Assume that (C) holds.
Take any $\vc{x}_1, \vc{x}_2 \in {\mathcal{X}}^N$ such that $\Phi\vc{x}_1=\Phi\vc{x}_2$.
Since $\vc{x}_1-\vc{x}_2\in\tilde{\mathcal{X}}^N$ and $\Phi$ is injective on $\tilde{\mathcal{X}}^N$,
we have $\vc{x}_1-\vc{x}_2=\vc{0}$, or $\vc{x}_1=\vc{x}_2$.
\end{IEEEproof}

Throughout the letter, we assume the uniqueness of the solution,
that is, the pair $({\mathcal{X}},\Phi)$ is chosen to satisfy
\eqref{eq:kerP}.
If the uniqueness assumption holds, we can find the exact solution in a finite number of steps
via an exhaustive computation as follows.
The set ${\mathcal{X}}^N$ is a finite set,
and we can write $\mathcal{X}^N=\{\vc{x}_1,\vc{x}_2,\ldots,\vc{x}_\mu\}$.
For each $\vc{x}_i$, we compute
$\vc{y}_i = \Phi \vc{x}_i$
and check if $\vc{y}_i=\vc{y}$.
Thanks to the uniqueness assumption, we can find the exact solution
in a finite time.
The problem is that the size of ${\mathcal{X}}^N$ is $\mu=L^N$,
and hence the computational complexity is exponential.
For example, if $L=2$ (two symbols) and $N=200$,
then 
\[
\mu=2^{200}\approx1.7\times10^{60},
\]
which takes at worst about $1.5\times10^{36}$ years 
(much longer than the lifetime of the universe) by the
current fastest computer with $34$ peta FLOPS.
To overcome this, we adopt a relaxation technique
based on the sum of absolute values in the next section.

\section{Solution via Sum of Absolute Values}
\label{sec:solution}
\begin{figure}[tb]
\centering
\includegraphics[width=0.9\linewidth]{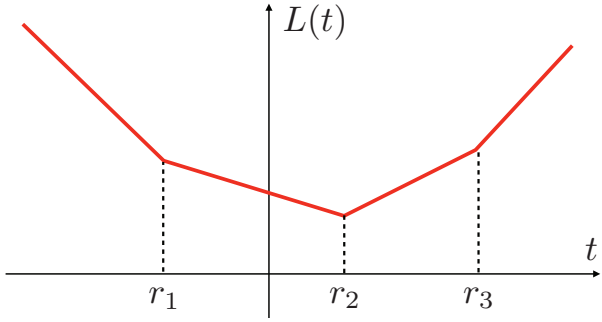}
\caption{Piecewise linear function $L(t)$ in the cost function \eqref{eq:Jx}.}
\label{fig:Lt}
\end{figure}

We here propose a relaxation method for discrete signal reconstruction.
By borrowing the idea of compressed sensing,
we can assume that each vector $\vc{x}-r_i$, $i=1,2,\ldots,L$, is sparse
(if the probability $p_i$ given in \eqref{eq:Prob} is not so small),
and the sparsity is proportional to the probability $p_i$.
Hence, we consider the following minimization problem:
\begin{equation}
 \underset{\vc{z}}{\mathrm{minimize}}~ F(\vc{z})\triangleq\sum_{i=1}^L p_i \| \vc{z}-r_i\|_1~\mathrm{subject~to}~ \vc{y}=\Phi\vc{z},
 \label{eq:P1}
\end{equation}
where we use the $\ell^1$ norm for the measure of sparsity as in compressed sensing.
By the definition of the $\ell^1$ norm,
we rewrite the cost function $F(\vc{z})$ as
\begin{equation}
  F(\vc{z}) = \sum_{i=1}^L p_i \sum_{n=1}^N |z_n-r_i| = \sum_{n=1}^N L(z_n),
  \label{eq:Jx}
\end{equation}
where $L(t)$ is the sum of weighted absolute values
\[
 L(t) \triangleq \sum_{i=1}^L p_i|t-r_i|.
\]
An example of this function is shown in Fig.~\ref{fig:Lt}.
As is shown in this figure,
the function $L(t)$ is continuous, convex,
and piecewise linear.
In fact, we have the following proposition:
\begin{proposition}
The function $L(t)$ is continuous and convex on ${\mathbb R}$ and is a
piecewise linear function given by
\[
 L(t) = \begin{cases}
 		-t + \overline{r}, &\text{if~} t \in (-\infty,r_1]\\
		a_it + b_i,  &\text{if~} t \in (r_i, r_{i+1}],~ i=1,2,\ldots,L-1\\
		t-\overline{r}, &\text{if~} t \in [r_L,\infty).
	\end{cases}
\]
where
\begin{gather*}
  \overline{r} \triangleq \sum_{i=1}^L p_ir_i,\\
  a_i \triangleq \sum_{j=1}^i p_j -\sum_{j=i+1}^L p_j,~~
  b_i \triangleq -\sum_{j=1}^i p_jr_j +\sum_{j=i+1}^L p_jr_j.
\end{gather*}
\end{proposition}
\begin{IEEEproof}
Since each function $|t-r_i|$ in $L(t)$ is continuous and convex
on $\mathbb{R}$,
the function $L(t)$, which is the convex combination of $|t-r_i|$, $i=1,2,\ldots,L$,
is also continuous and convex on $\mathbb{R}$.

Suppose $t\leq r_1$. From the inequality \eqref{eq:rl},
we have $t - r_i \leq 0$ for $i=1,2,\ldots,L$, and hence
\[
 L(t) = -\sum_{i=1}^L p_i(t-r_i) = -t + \overline{r},
\]
where we used \eqref{eq:p}.
Next, suppose $r_i < t \leq r_{i+1}$ ($i=1,2,\ldots,L-1$).
The inequality \eqref{eq:rl} gives
$t - r_j > 0$ for $j=1,2,\ldots,i$
and $t - r_j \leq 0$ for $j=i+1,i+2,\ldots,L$. It follows that
\[
 L(t) = \sum_{j=1}^i p_j(t-r_j) - \sum_{j=i+1}^L p_j(t-r_j)
  = a_i t + b_i.
\]
Finally, if $t\geq r_L$, then $t-r_i \geq 0$ for $i=1,2,\ldots,L$ due to \eqref{eq:rl},
and hence
\[
 L(t) = \sum_{i=1}^L p_i(t-r_i) = t - \overline{r}.
\] 
\end{IEEEproof}

Note that the function $L(t)$ is rewritten as
\[
 L(t) = \max_{i=0,1,\ldots,L} \{a_i t + b_i\},
\]
where $a_0=-1$, $b_0=\overline{r}$, $a_L=1$, and $b_L=-\overline{r}$.
It follows that the optimization \eqref{eq:P1} is equivalently described as
\begin{equation}
\begin{aligned}
& \underset{\vc{\theta}\in{\mathbb{R}}^N}{\text{minimize}}
 & & \vc{1}_N^\top \vc{\theta}\\
& \text{subject to}
& & \vc{y}=\Phi\vc{z},
& A\vc{z}+\vc{b} \leq E\vc{\theta},
\end{aligned}
\label{eq:LP}
\end{equation}
where $\vc{\theta}\in{\mathbb{R}}^N$ is
an auxiliary variable, and
\[
 \begin{split}
 A &\triangleq I_N \otimes \vc{a}\in{\mathbb{R}}^{N(L+1)\times N},
  \vc{a}\triangleq [a_0,a_1,\ldots,a_L]^\top\in{\mathbb{R}}^{L+1},\\
 \vc{b} & \triangleq \vc{1}_N\otimes\vc{b}\in{\mathbb{R}}^{N(L+1)},
  \vc{b}\triangleq [b_0,b_1,\ldots,b_L]^\top \in{\mathbb{R}}^{L+1},\\
  E &\triangleq I_N \otimes \vc{1}_{L+1}\in {\mathbb{R}}^{N(L+1)\times N}.
 \end{split} 
\] 
This is a standard linear programming problem and can be efficiently solved
by numerical optimization softwares, such as {\tt cvx} in 
{\tt MATLAB}~\cite{cvx,GraBoy08}.

Now, we discuss the validity of the relaxation optimization
given in \eqref{eq:P1} or \eqref{eq:LP}.
To see this, we extend the notion of the {\em null space property}~\cite{CohDahDeV09}
in compressed sensing to our problem:
\begin{definition}
A matrix $\Phi\in{\mathbb{C}}^{M\times N}$ is said to satisfy
the {\em null space property} for an alphabet $\mathcal{X}$ if
\begin{equation}
 F(\vc{x}) < F(\vc{x}-\vc{v}),
\end{equation}
for any $\vc{x}\in{\mathcal{X}}^N$ and any $\vc{v}\in {\mathrm{ker}}\,\Phi\setminus\{\vc{0}\}$.
\end{definition}
Then we have the following theorem:
\begin{theorem}
Let $\Phi\in{\mathbb{C}}^{M\times N}$.
Every $\vc{x}\in{\mathcal X}^N$ is uniquely recovered from
the $\ell^1$ optimization \eqref{eq:P1} with $\vc{y}=\Phi\vc{x}$
if and only if $\Phi$ satisfies the null space property for $\mathcal{X}$.
\end{theorem}
\begin{IEEEproof}
($\Rightarrow$):
Take any $\vc{v}\in{\mathrm{ker}}~\Phi\setminus\{0\}$ and $\vc{x}\in{\mathcal{X}}^N$.
Put $\vc{z}\triangleq \vc{x}-\vc{v}$.
Since $\vc{v}\in\mathrm{ker}\,\Phi$, we have
$\Phi\vc{v}=\Phi(\vc{v}-\vc{x}+\vc{x})=\vc{0}$,
or
$\Phi\vc{x}=\Phi\vc{z}$.
This means that $\vc{z}$ is in the feasible set of the optimization problem \eqref{eq:P1}
with $\vc{y}=\Phi\vc{x}$.
Also, we have $\vc{x}\neq\vc{z}$ since $\vc{v}\neq\vc{0}$.
Now, by assumption, $\vc{x}$ is the unique solution of \eqref{eq:P1} with $\vc{y}=\Phi\vc{x}$,
and hence we have
$F(\vc{x})<F(\vc{z})=F(\vc{x}-\vc{v})$.

($\Leftarrow$):
Take any $\vc{x}\in{\mathcal{X}}^N$ and $\vc{z}\in{\mathbb{C}}^N$ such that
$\vc{x}\neq\vc{z}$ and $\Phi\vc{x}=\Phi\vc{z}$.
Put $\vc{v}\triangleq\vc{x}-\vc{z}$.
Then we have $\vc{v}\in\mathrm{ker}\,\Phi\setminus\{0\}$.
From the null space property for ${\mathcal{X}}$,
we have
$F(\vc{x})<F(\vc{x}-\vc{v})=F(\vc{z})$.
It follows that $\vc{x}$ is the unique solution of \eqref{eq:P1}.
\end{IEEEproof}

\section{Examples}
\label{sec:example}
In this section, we show two examples to illustrate the effectiveness of the proposed method.

The first example is one-dimensional signal reconstruction with multiple symbols.
Let the original signal $\vc{x}$ be a $200$-dimensional vector (i.e. $N=200$)
and the elements are drawn from the following alphabets
with probability distributions:
\begin{equation}
 \begin{split}
{\mathcal{X}}_2 &\triangleq \{0,1\}:  {\mathbb{P}}(0)=p,~{\mathbb{P}}(1)=1-p,\\
{\mathcal{X}}_3 &\triangleq \{-1,0,1\}:  {\mathbb{P}}(0)=p,~{\mathbb{P}}(1)={\mathbb{P}}(-1)=\frac{1-p}{2},\\
{\mathcal{X}}_5 &\triangleq \{-2,-1,0,1,2\}:  {\mathbb{P}}(0)=p,\\
 &\quad {\mathbb{P}}(-2)={\mathbb{P}}(-1)={\mathbb{P}}(1)={\mathbb{P}}(2)=\frac{1-p}{4},
\end{split}
 \label{eq:example_P}
\end{equation}
where $p\in[0,1]$.
We assume the measurement vector $\vc{y}$ is a $100$-dimensional vector (i.e. $M=100$), and
the measurement matrix $\Phi$ is generated such that
each element is independently drawn from 
the Gaussian distribution
with a mean of 0 and a standard deviation of $1$
by using a MATLAB command, {\tt randn(100,200)}.
The original vector $\vc{x}$ is also generated such that
each element is independently drawn from the distribution \eqref{eq:example_P}
with varying parameter $p\in[0,1]$.
\begin{figure}[tb]
\includegraphics[width=\linewidth]{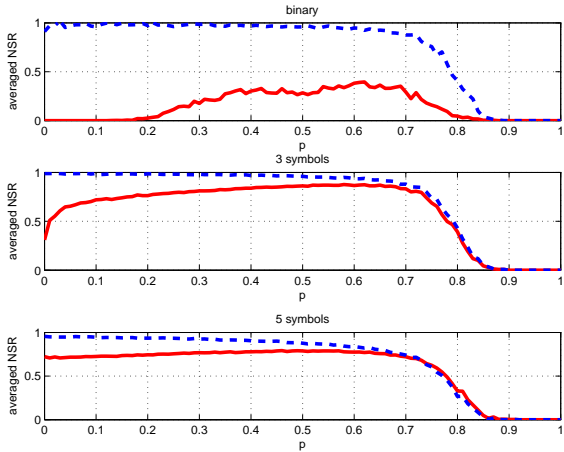}
\caption{Averaged NSR $\|\vc{x}-\hat{\vc{x}}\|_2/\|\vc{x}\|_2$ vs probability $p$
for $\mathcal{X}_2$ (top), $\mathcal{X}_3$ (middle), $\mathcal{X}_5$ (bottom),
by the proposed (solid) and the basis pursuit (dash)}
\label{fig:error}
\end{figure}
Fig.~\ref{fig:error} shows the graphs of
the averaged NSR (noise-to-signal ratio) $\|\vc{x}-\hat{\vc{x}}\|_2/\|\vc{x}\|_2$
for $\mathcal{X}_2,\mathcal{X}_3,\mathcal{X}_5$ in \eqref{eq:example_P}, 
where $\hat{\vc{x}}$ is the reconstructed signal,
with 200 trials of random $\Phi$ and $\vc{x}$ for each $p\in[0,1]$.
If $p$ is large (i.e., $p\approx 1$), then the original vector $\vc{x}$
is sparse, and we also reconstruct the original signal by the basis pursuit
\[
 \min_{\vc{z}\in{\mathbb{R}}^N} \|\vc{z}\|_1 \text{~subject to~} \vc{y}=\Phi\vc{z},
\]
and then round off the values by the basis pursuit to the nearest integer.
The error graphs by the basis pursuit are also shown in Fig.~\ref{fig:error}.
For the binary alphabet $\mathcal{X}_2=\{0,1\}$,
one may exchange the roles of 0 and 1 before performing the basis pursuit,
and the error curve below $p=0.5$ is pessimistic.
However, 
such a simple strategy cannot be applied
to $\mathcal{X}_3$ and $\mathcal{X}_5$ for the basis pursuit,
while the proposed method works well for small $p$.
This is because the basis pursuit does not fully utilize the information of
the alphabet (i.e., the basis pursuit only uses
the information of the value $0$ through the sparsity).
We also note that the basis pursuit can be used only when the alphabet includes
$0$, while the proposed method works as well when $0$ is not an element of the alphabet.
Fig.~\ref{fig:error} also implies a conjecture
that the performance of the proposed method
converges that of the basis pursuit as the size of the alphabet goes to infinity.

\begin{figure}[tb]
\centering
\includegraphics[width=0.48\linewidth]{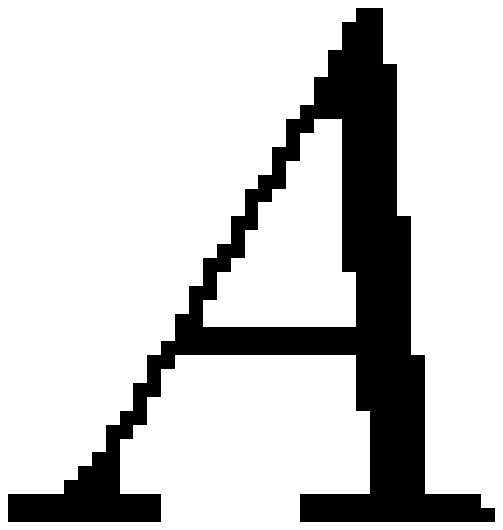}~
\includegraphics[width=0.48\linewidth]{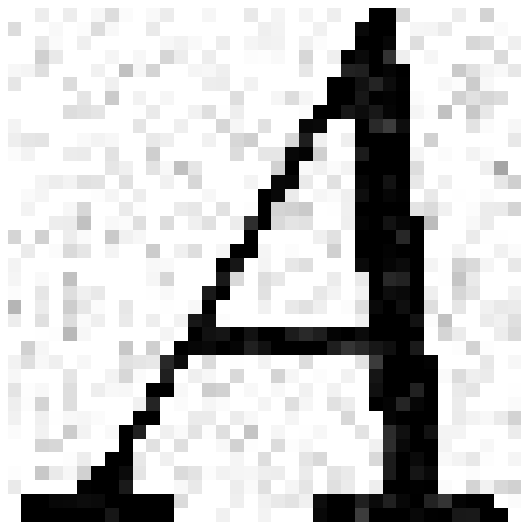}
\caption{Original image (left) and disturbed image by random noise (right)}
\label{fig:original_A}
\end{figure}

\begin{figure}[tb]
\centering
\includegraphics[width=0.48\linewidth]{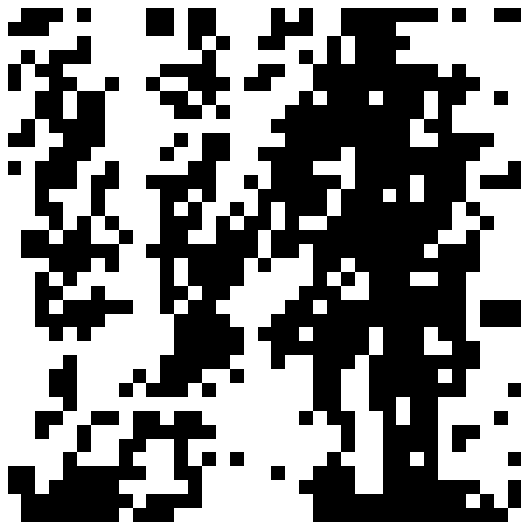}~
\includegraphics[width=0.48\linewidth]{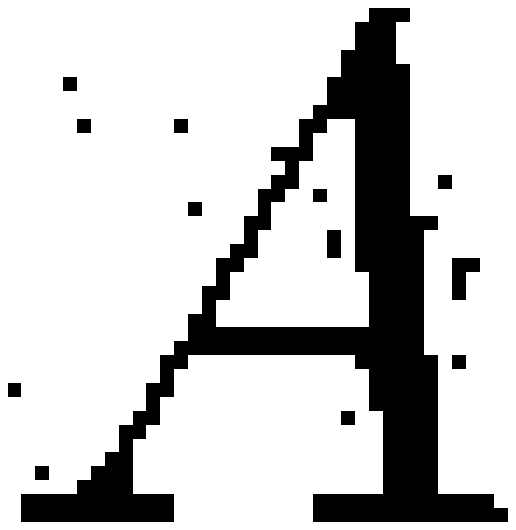}
\caption{Reconstructed images by the basis pursuit (left) and by the proposed method (right)}
\label{fig:reconstruction_A}
\end{figure}

Next, we see an example from image processing.
Let us consider a binary (or black-and-white) image shown in Fig.~\ref{fig:original_A} (left),
which is a $37\times 37$-pixel binary-valued image.
We add random Gaussian noise with a mean of 0 and a standard deviation of $0.1$ to each pixel
to obtain a disturbed image as shown in Fig.~\ref{fig:original_A} (right).
We represent this disturbed image as a real-valued matrix $X\in {\mathbb{R}}^{37\times 37}$.
Then we apply the discrete Fourier transform (DFT) to $X$ to obtain
\begin{equation}
 \hat{X} = WXW \label{eq:DFT}
\end{equation}
where $W$ is the DFT matrix defined by
\[
 W \triangleq 
  \begin{bmatrix}
   1&1&1&\ldots&1\\
   1&\omega&\omega^2&\ldots&\omega^{K-1}\\
   \vdots&\vdots&\vdots&\ddots&\vdots\\
   1&\omega^{K-1}&\omega^{2(K-1)}&\ldots&\omega^{(K-1)(K-1)}
  \end{bmatrix}, 
\]
where $K=37$ and $\omega\triangleq \exp(-\mathrm{j}2\pi/K)$.
The relation can be equivalently represented by
\[
 \mathrm{vec}(\hat{X}) = (W\otimes W)\mathrm{vec}(X) \in \mathbb{C}^{1369}. 
\]
We then randomly down-sample the vector $\mathrm{vec{(\hat{X})}}$
to obtain a half-sized vector $\vc{y}\in\mathbb{C}^{685}$.
The measurement matrix $\Phi$ is then a $685\times 1369$ matrix
generated by randomly down-sampling row vectors from $W\otimes W$.
Fig.~\ref{fig:reconstruction_A} shows the reconstructed images
by the basis pursuit with rounding off (left) and
by the proposed method (right).
For the proposed method, we assumed $\mathbb{P}(0)=\mathbb{P}(1)=1/2$.
The results clearly show the effectiveness of our method also for image reconstruction.


\section{Conclusion}
\label{sec:conclusion}
In this letter, we have proposed a reconstruction method for discrete signals
based on the sum of absolute values (or the weighted $\ell^1$ norm).
The reconstruction algorithm is described as linear programming, which 
can be solved effectively by numerical optimization softwares.
Examples have been shown that the proposed method is much more effective
than the basis pursuit which only uses the information of sparsity.
Future work includes an accessible condition that ensures
the null space property, as the restricted isometry property in compressed sensing.

\section*{Acknowledgment}
This research is supported in part by the JSPS Grant-in-Aid for Scientific Research (C) No.~24560543,
Grant-in-Aid for Scientific  Research on Innovative Areas
No.~26120521,
and an Okawa Foundation Research Grant.

\ifCLASSOPTIONcaptionsoff
  \newpage
\fi

\newpage



%


\end{document}